\documentclass{MyArticle}

\begin{document}

\begin{center}
\begin{minipage}{.8\textwidth}
	\begin{center}
	{\Huge Thermal time of noncommutative Minkowski spacetime}
	
	\vspace{20pt}
	
	{\Large Kilian Hersent}\footnote{%
		\ \href{mailto:khersent@ubu.es}{khersent@ubu.es}
	}
	
	\vspace{7pt}	
	
	{\large Departamento de F\'{i}sica, Universidad de Burgos, 09001 Burgos, Spain.}
	
	\vspace{15pt}
	
	{\textbf{Abstract}}
	
	\end{center}
	
	\paragraph{}
	In this paper, we study the thermal time hypothesis of \cite{Connes_1994a} in the context of noncommutative deformations of Minkowski. We show that a natural modular group arises from the modular function of the momentum space. In the specific case of $\kappa$-Minkowski, we show that this thermal time flow corresponds to the globally defined time coordinate translation. On the other hand, the absence of thermal time for $\rho$-Minkowski is directly related to the discreteness of its global time. The impact of inner automorphism transformation on the physics and the treatment of unimodular case (unthermalised spacetimes) are discussed. Moreover, a reflection on the use of thermal field theory for quantum gravity phenomenology is put forward, as we just bridged thermal spacetimes with $\kappa$-Minkowski, often considered a ``flat limit'' of a quantum gravity candidate theory.
\end{minipage}
\end{center}

\tableofcontents

\section{Introduction}
\paragraph{}
The gap between the treatment of quantum field theory and general relativity generates theoretical issues in the construction of a quantum theory of gravity. One of the major issue is known as the ``problem of time'' \cite{Anderson_2012} and emerges from an incompatible treatment of the time in both theories. In quantum field theory, the time is used to parametrise the dynamics of all observables as $A(t) \vert\psi\rangle = e^{itH} A e^{-itH} \vert\psi\rangle$, where $H$ is the Hamiltonian. Even in a relativistic context, such a parametrisation can be made sense of since there exists a covariant notion of time, the proper time. In general relativity though, no generally covariant notion of time exists so that space and time are fully ``merged'' into spacetime. An evolution defined in the sense above would necessarily trigger a notion of (dynamical) time that is observer dependant.

\paragraph{}
As a tentative proposal to bridge this gap, the authors of \cite{Rovelli_1993a, Rovelli_1993b, Connes_1994a} have tried to built a notion of time that could be defined in a generally covariant context. Their idea ends up considering the algebraic tools of noncommutative geometry to define a thermal time associated to any state of the theory. However, compared to our previous analysis, this notion of thermal time changes from one state to the other in a controlled way, so that even in a generally covariant context, this thermal time could parametrise the dynamics of the state. They even formulate the ``thermal time hypothesis'' which states that this notion of thermal time corresponds to the physical time of the associated state.

This definition of time has allowed advances in algebraic quantum field theory \cite{Borchers_2000} and a thermodynamic formulation of general relativity \cite{Vidotto_2024}. Yet, the establishment of this time in the context of a quantum geometry has never been done, even after 30 years. We propose in this paper to construct the first thermal time \emph{of} a noncommutative spacetime, by considering deformations of the Minkowski space. 

It is very important to note that we are here performing the thermal time analysis \textit{of} the (noncommutative) spacetime and not of a (quantum) observer \textit{on} the spacetime, as done for a quantum gas or an accelerated observer in \cite{Connes_1994a}. The difference relies in the considered algebra to construct the thermal time. In our case, the algebra is the (deformed) algebra of smooth functions on the geometry, which in a noncommutative geometry context defines the geometry.

\paragraph{}
In order to define the thermal time on the deformed Minkowski spacetime, we intensively use the properties of its momentum space. The modular operator, generating the dynamics, is shown to be closely related to the modular function of the momentum space. The emergence of a thermal behaviour from noncommutativity is commented throughout the paper and could open the door to a new range of quantum gravity phenomenology.

The explicit case of $\kappa$-Minkowski is derived and it is shown that the thermal time evolution corresponds to a (global) time translation. This result is quite promising for the thermal time hypothesis since both notions of time (thermal and geometrical) coincide in this context. Note that the global time is however not covariant with respect to (deformed) Poincar\'{e} transformations, implying that the thermal time is actually selecting a preferred frame. This result is in full accordance with thermal field theory, for which the preferred frame is the equilibrium one.

The case of $\rho$-Minkowski is also analysed. As it consists of a unimodular spacetime, the thermal time cannot be define for the tracial state. However, it is explicitly shown that the unimodularity property can be directly related to the discreteness of the time (spectrum) for this spacetime. It is quite coherent with the fact that the thermal time defines a continuous time flow, which is not compatible with a discrete global time.

Finally, the paper also tries to question the inner automorphism transformations, which corresponds to a transformation of the time flow after changing of equilibrium state. The impact of such a transformation on the physics of the thermal time has never been questioned, as far as the author knows, and may have a considerable impact on the physics.

\paragraph{}
This paper is organised as follows. The construction of a noncommutative deformation of Minkowski is detailed in section \ref{sec:Md}. The section \ref{sec:tt} builds the notion of thermal time in a general context and in the case of the deformed Minkowski space. It also discusses open problems concerning the thermal time. The explicit examples of $\kappa$-Minkowski and $\rho$-Minkowski are treated in section \ref{sec:kM}. The concluding section \ref{sec:c} gathers the main results together with an open discussion on quantum gravity phenomenology and other unsolved problems. The appendix \ref{apx:KMS} gathers some useful proofs. Throughout the paper, the convention of summation over repeated indices is applied. We also work in natural units $c = \hbar = k_B = 1$.

\section{Deformations of Minkowski and its momentum space}
\label{sec:Md}
\paragraph{}
In this section, we give a technical overview of how noncommutative deformations of Minkiwski are built. Many details are skipped but can be found in the relevant references. It gathers known results but written down in a quite general formulation and consists of an unprecedented effort of generalisation, as far as the author knows.

\subsection{Quantum deformation of Poincar\'{e}}
\label{subsec:Md_Pd}
\paragraph{}
We consider here a noncommutative deformation of the Minkowski space in the Hopf algebra formalism. This kind of approach is based on the construction of a quantum deformation, in the sense of quantum groups, of the Poincar\'{e} algebra, and then defines the deformed Minkowski space as the Hopf algebra dual to the translation Hopf subalgebra of the deformed Poincar\'{e}. This approach was first use to define the $\kappa$-Minkowski space in \cite{Majid_1994}, but was also used to build other deformations of Minkowski (or $\Real^4$), such as the Moyal space (sometimes called $\theta$-Minkowski)\footnote{
	Before any confusion arises, the Moyal space was not built through a deformation of Poincar\'{e} at first, and the symmetries of Moyal led to an extensive debate, as mentioned in \cite{Mercati_2023} (see references therein). However, it is now known that one can construct Moyal through a deformation of Poincar\'{e}, dubbed $\theta$-Poincar\'{e}, for example via a Drinfel'd twist \cite{Aschieri_2005}.
}
or the $\rho$-Minkowski space \cite{Dimitrijevic_2018}. For a quite complete and descriptive introduction to those deformations, see the introduction of \cite{Mercati_2023}.

\paragraph{}
Let us consider such a Hopf algebra deformation of the $d+1$-dimensional Poincar\'{e}, named $\Poin{d}_\ell$ hereafter, without prescribing how it has been constructed. We indeed want to keep some generality here and show that our computations do not depend on a specific deformation. We call $\{P_\mu\}_{\mu=0, \ldots, d}$ the generator of the (deformed) infinitesimal translation and make the hypothesis that they form a Hopf subalgebra of $\Poin{d}_\ell$, noted $\Tran{d}_\ell$. The deformed Minkowski space, noted $\mathbb{A}_\ell$ hereafter\footnote{
	This notation is due to the fact that we are considering a noncommutative deformation of the \emph{algebra} of smooth functions on Minkowski, \ie of $C^\infty(\Mink{d})$, which thus forms a noncommutative algebra $\mathbb{A}_\ell$. This idea is at the very basis of noncommutative geometry \cite{Connes_1994b}. The author has proposed a tentative introduction to the essential ideas of noncommutative geometry in Section 1 of \cite{Hersent_2024b}.
},
is defined as the Hopf dual to $\Tran{d}_\ell$, where $\{x^\mu\}_{\mu = 0, \ldots, d}$ is the dual basis of $\{ P_\mu \}_{\mu = 0, \ldots, d}$. By denoting the dual pairing as $\langle \cdot, \cdot \rangle : \Tran{d}_\ell \times \mathbb{A}_\ell \to \Cpx$, the dual basis condition writes $\langle P_\mu, x^\nu \rangle = \delta_\mu^\nu$.

First, note that the duality fully determines the Hopf algebra structure of $\mathbb{A}_\ell$. Furthermore, as Hopf algebras, the elements of $\Tran{d}_\ell$ and $\mathbb{A}_\ell$, can be written as (possibly infinite) polynomials of the generators, that is
\begin{align}
	f
	&= \sum_{n=0}^{+\infty} f_{\mu_1, \ldots, \mu_n} x^{\mu_1} \cdots x^{\mu_n}, &
	\hat{f}
	&= \sum_{m = 0}^{+\infty} \hat{f}^{\nu_1, \ldots, \nu_m} P_{\nu_1} \cdots P_{\nu_m},
	\label{eq:Md_gen}
\end{align}
where $f_{\mu_1, \ldots, \mu_n}, \hat{f}^{\nu_1, \ldots, \nu_m} \in \Cpx$, for any $f \in \mathbb{A}_\ell$ and any $\hat{f} \in \Tran{d}_\ell$. In this respect, the dual pairing extends to
\begin{align}
	\big\langle P_{\nu_1} \cdots P_{\nu_m}, x^{\mu_1} \cdots x^{\mu_n} \big\rangle
	&= \delta_{n,m} \delta^{\mu_1}_{\nu_1} \cdots \delta^{\mu_n}_{\nu_n} &
	& \Longrightarrow &
	\langle \hat{f}, f \rangle
	&= \sum_{n=0}^{+\infty} \hat{f}^{\mu_1, \ldots, \mu_n} f_{\mu_1, \ldots, \mu_n}.
	\label{eq:Md_dp}
\end{align}

\paragraph{}
It is important to note that the deformation parameter, noted here $\ell$, is of primordial physical interest. Mathematically speaking, it measures ``how much'' the quantity we are considering depart from the undeformed quantities. But, in most deformations (like $\theta$, $\kappa$ and $\rho$-deformations), $\ell$ has a dimension, turning the deformation parameter into a scale intrinsic to the quantum space\footnote{
	As examples, $\theta$ is a surface (like $\kappa^{-2}$ and $\rho^2$), often associated to the minimal observable surface of the deformed spacetime thanks to Heisenberg uncertainties on coordinates. We refer to the Doplicher-Fredenhagen-Roberts framework \cite{Doplicher_1995} for a physical interpretation of these uncertainties. $\kappa^{-1}$ and $\rho$ (and $\sqrt{\theta}$) are lengths, thought to have some relation with the Planck length.
}.
One of the new feature of this paper is that the deformation parameter may parametrise a (high) temperature scale \emph{of} the spacetime. In the example of $\kappa$-Minkowski, developed in section \ref{sec:kM}, $\kappa$ is associated to the equilibrium temperature of the tracial weight.

\subsection{From symmetries to position and momenta}
\label{subsec:Md_pm}
\paragraph{}
The duality between $\Tran{d}_\ell$ and $\mathbb{A}_\ell$ allows us to define a $\Tran{d}_\ell$-module algebra structure on $\mathbb{A}_\ell$ (see for example Section 1.3.5 of \cite{Klimyk_1997}). It corresponds to the structure relating a quantum space and its symmetries\footnote{
	Not having found explicit explanations on this topic, the author has tried to make his own understanding, displayed in Appendix A.2 of \cite{Hersent_2024b}.
}.
Explicitly, one can define an action $\actl : \Tran{d}_\ell \otimes \mathbb{A}_\ell \to \mathbb{A}_\ell$ of the (infinitesimal) deformed translations on the deformed algebra of functions on the spacetime. It is defined as being linear, compatible with the product of $\Tran{d}_\ell$ and linking the product and unit of $\mathbb{A}_\ell$ to the coproduct and counit of $\Tran{d}_\ell$ (respectively). For an explicit definition, we refer to \cite{Klimyk_1997, Hersent_2024b} out of many other references.

\paragraph{}
We now define the position and momentum space as being the elements of $\mathbb{A}_\ell$ and $\Tran{d}_\ell$ respectively, that are linear in the generators. Explicitly, the elements for which the decomposition \eqref{eq:Md_gen} writes
\begin{align*}
	x
	&= a_\mu x^\mu, &
	p
	&= v^\mu p_\mu
\end{align*}
for $a_\mu, v^\mu \in \Cpx$, $x \in \mathbb{A}_\ell$ and $p \in \Tran{d}_\ell$. Note that we have changed the notation of the momentum space generators to lower-case letter $\{p_\mu\}_{\mu=0,\ldots,d}$ as in the physics literature. The position space is denoted $\Mink{d}_\ell$, with generic element $x$ as above. The momentum space is noted $G_\ell$, with generic element $p$. Note that in general $(\Mink{d}_\ell, +, \cdot)$ and $(G_\ell, + , \cdot)$ do not form (Lie)-algebras since they may not be closed by product (or bracket).

Let us consider a plane wave that we denote generically by $e^{i p x}$. The product between $p$ and $x$ is here formal and can be chosen to be a tensor product $\otimes$. We omit it throughout the paper for simplicity. There are several ways to define a plane wave, as for examples
\begin{subequations}
\begin{align}
	e^{i p x}
	&= e^{i (p_0 x^0 + \cdots + p_d x^d)}
	= \sum_{n = 0}^{+ \infty} \frac{i^n}{n!} p_{\mu_1} \cdots p_{\mu_n} x^{\mu_1} \cdots x^{\mu_n}, \\
	e^{i p x}
	&= e^{i p_0 x^0} \cdots e^{i p_d x^d}
	= \left( \sum_{n_0=0}^{+\infty} \frac{i^{n_0}}{n_0!} (p_0)^{n_0} (x^0)^{n_0} \right) \cdots \left( \sum_{n_d=0}^{+\infty} \frac{i^{n_d}}{n_d!} (p_d)^{n_d} (x^d)^{n_d} \right) \\
	e^{i p x}
	&= e^{i p_d x^d} \cdots e^{i p_0 x^0}
	= \left( \sum_{n_d=0}^{+\infty} \frac{i^{n_d}}{n_d!} (p_d)^{n_d} (x^d)^{n_d} \right) \cdots \left( \sum_{n_0=0}^{+\infty} \frac{i^{n_0}}{n_0!} (p_0)^{n_0} (x^0)^{n_0} \right)
\end{align}
\end{subequations}
among many other possibilities. In the commutative limit, all those expressions converge to the same one, \ie the plane wave on the classical spacetime. As those orderings are indistinguishable and that there are no argument in favour of choosing one more then another, there is an ordering ambiguity of noncommutative plane waves, which has been pointed out in \cite{Mercati_2018} (see also \cite{Hersent_2024a} for an extended discussion on such an ambiguity for noncommutative quantum field theory).

Such plane waves are useful to gather information on the momentum space. Indeed, if one choose an ordering\footnote{
	Note that an ordering is only necessary if $[P_\mu, P_\nu] \neq 0$, that is if the momentum space is noncommutative. In most known deformations ($\theta$, $\rho$, $\kappa$, \textit{etc}...) the momentum space is commutative.
}
adapted to the action $\actl$ defined above, one has that
\begin{align}
	P_\mu \actl e^{i p x} = p_\mu e^{i p x}.
	\label{eq:Md_Pev}
\end{align}
One could think at this point that the chosen ordering is fixed once and for all, which means we have lost some degrees of freedom in the theory. However, one can go from one ordering to another by changing variable in the $p$'s and so conserve \eqref{eq:Md_Pev} by performing a similar change of variable on the $P$'s. This was proven in the specific case of $\kappa$-Minkowski in \cite{Mercati_2018}. The ordering ``degree of freedom'' has thus been transferred to the coordinate choice invariance on the momentum space.

\paragraph{}
If one considers the product of two plane-waves, one can write, using the definition of the action $\actl$,
\begin{align*}
	P_\mu \actl (e^{i p x} e^{i q x})
	&= \big(P_{\mu(1)} \actl e^{i p x} \big) \big( P_{\mu(2)} \actl e^{i q x} \big) \\
	&= p_{\mu(1)} q_{\mu(2)} (e^{i p x} e^{i q x}) \\
	&:= (p \dplus q)_\mu (e^{i p x} e^{i q x}) \\
	&= P_\mu \actl e^{i (p \dplus q) x},
\end{align*}
where $\Delta(P_\mu) = P_{\mu(1)} \otimes P_{\mu(2)}$ is the Sweedler notations for the coproduct and we have defined a new law in $G_\ell$, named $\dplus$, using this coproduct. We have also showed here that the product of two plane waves can again be written as a plane wave involving this new law $\dplus$. One can show (as done in \cite{Hersent_2024a}) that $(G_\ell, \dplus)$ forms a group, with unit $0$ and inverse law $\dminus$, defined through the counit and antipode of $\Tran{d}_\ell$ respectively. It can also straightforwardly be shown that in the commutative limit (here $\ell \to 0$), the $\dplus$ law corresponds to the usual addition law $+$ of momenta. It is thus frequently described as a deformed (or noncommutative) addition law for momenta. This law is noncommutative if and only if the spacetime coordinates $x^\mu$ of $\Mink{d}_\ell$ do not commute among one another, as indeed both premisses are equivalent to $\Tran{d}_\ell$ being non-cocommutative.

\subsection{Properties of the momentum space}
\paragraph{}
In order to go further, we need to characterise the momentum space $G_\ell$. First, let us state that $(G_\ell, \dplus)$ viewed as a group, is locally compact at least for the $\ell$-adic topology. This means that the development we are going to make here is true at least when considering our objects as formal power series in $\ell$, the deformation parameter\footnote{
	For a pedagogical introduction to $\hbar$-adic topology in the context of deformation quantisation, see for example the Appendix A of \cite{Pachol_2011}.
}.

Note that it could be proven that $(G_\ell, \dplus)$ is locally compact in other cases, like when the $x$'s form a Lie algebra, \ie $[x, x] \propto x$. In this case, $(G_\ell, \dplus)$ corresponds to the $d+1$-dimensional Lie group of this Lie algebra, which is automatically locally compact. The proof of the local compactness in the general case is necessary for what we do here, but go far beyond the scope of this paper.

\paragraph{}
The fact that $G_\ell$ is locally compact allow us to define a Haar measure on $(G_\ell, \dplus)$ and so to perform integration over the momentum space. This is developed in details in \cite{Hersent_2024a}. For the mathematical aspects of harmonic analysis, we refer to textbooks like \cite{Deitmar_2014}. In general, there are two distinct Haar measures: the left invariant one, noted here $\Haar{p}$, for $p \in G_\ell$ and the right invariant one. In the following, we only make use of the left Haar measure, but one could equivalently work with the right invariant one. The ratio between these two measures is known as the modular function of the group, noted $\Delta : G_\ell \to \spReal$. It is a strictly positive continuous morphism. In order to avoid confusion in notations with the coproduct, we will rather consider the inverse modular function, noted $\mathscr{I}(p) = \Delta(\dminus p) = 1 / \Delta(p)$ for any $p \in G_\ell$, and only refer to it in the computations.

In the specific case where $\mathscr{I} = 1$, the group $G_\ell$ is said to be unimodular and the left and right Haar measures are in fact equal. Note that the definition of the thermal time of $\Mink{d}_\ell$, given in the following, heavily relies on the non-trivial structure of this momentum group $(G_\ell, \dplus)$ and the fact that it is not unimodular, as discussed below.

\subsection{The \tops{$C^*$}{C*}-algebra structure}
\label{subsec:Md_Cs}
\paragraph{}
Before handling the notions used to define the thermal time, we need to give to $\mathbb{A}_\ell$ a $C^*$-algebra structure. We do not want to explicitly choose a $C^*$ structure here to keep some generality, but it needs to be proven that such a structure may exist. This construction generally go through the construction of a product (generically noted $\star$ and called the star-product) and an involution on $\mathbb{A}_\ell$. Those explicit constructions are used to extend the space of functions that defines $\mathbb{A}_\ell$ from polynomials (see \eqref{eq:Md_dp}) to a specific set of functions (corresponding to an extention of the compactly supported functions), that are sometimes referred to as the multiplier space.

This can be done by using generalised Weyl systems as introduced in \cite{Agostini_2002}. In order to obtain an explicit expressions of the product of $\mathbb{A}_\ell$ and its involution, that we note ${}^\dagger$, one can also use convolution algebra techniques, as developed for example in \cite{Durhuus_2013} (for a review see \cite{Hersent_2023a}). The latter approach uses the expression of $\dplus$, that one can directly obtain from the coproduct of $\Tran{d}_\ell$, to define a convolution algebra on $(G_\ell, \dplus)$ and, through it, a quantisation map. Another widely used method to obtain star-products consists of the use of Drinfel'd twist type of deformation\footnote{
	A pedagogical introduction to this topic was made in section 1.6.2 of \cite{Hersent_2024b}, but we refer more advanced readers to \cite{Aschieri_2014}.
}.
In any of those cases, the $*$-algebra structure is engineered  so that $\mathbb{A}_\ell$ fulfils the same Hopf algebra structure as obtained from dualising $\Tran{d}_\ell$. As already put forward, the explicit construction of this $*$-algebra structure in the general case goes far beyond the scope of this paper, but we wanted to underline that it exists.

\paragraph{}
Once the space of functions on $\mathbb{A}_\ell$ is obtained, it remains to have a norm and show that this norm fulfills the $C^*$-algebra axioms. Before discussing this point, we need to introduce the notion of Fourier transform and of delta functions. Such preliminaries actually allow us to bridge position space and momentum space and so to export notions and properties from the momentum space to the position space.

In order to do so, we consider here that the integral over Minkowski space $\Mink{d}$ can be exported to its deformed version, and forms a weight, we denote by $\omega : \mathbb{A}_\ell \to \Cpx$, \ie
\begin{align}
	\omega(f)
	&= \int \tdl{x} f(x)
	\label{eq:Md_w}
\end{align}
for any $f \in \mathbb{A}_\ell$. This can be done at least formally, by writing $f$ as a power series in $\ell$. Now that integration measures on $\mathbb{A}_\ell$ and $L^1(G_\ell)$ are built, we can define a Fourier transform and its inverse as
\begin{subequations}
	\label{eq:Md_Ft}
	\begin{align}
		\mathcal{F}(f)(p)
		&= \frac{1}{(2\pi)^{d+1}} \int \tdl{x} e^{i (\dminus p) x} f(x), \\
		\mathcal{F}^{-1}(\hat{f})(x)
		&= \int \Haar{p} e^{i p x} \hat{f}(p),
	\end{align}
\end{subequations}
where $f \in \mathbb{A}_\ell$ and\footnote{
	The hat is omitted in the following so that $f$ denotes both the function and its Fourier transform.
}
$\hat{f} \in L^1(G_\ell)$. It is a matter of algebra to show that $\mathcal{F} \circ \mathcal{F}^{-1} = \id$. Furthermore, these maps converges to the usual Fourier transform in the commutative limit. 

It is quite remarkable to see that even if the star-product has not been built through the convolution product of $(G_\ell, \dplus)$, the Fourier transform defined by \eqref{eq:Md_Ft} actually imposes that the momentum counterpart of the star-product is the convolution product \cite{Agostini_2002}. Explicitly, if one considers $f, g \in \mathbb{A}_\ell$,
\begin{align*}
	(fg)(x)
	&= \int \Haar{p} \Haar{q} f(p) g(q) e^{i(p \dplus q)x}
	\overset{(q \to \dminus p \dplus q)}{=} \int \Haar{p} \Haar{q} f(p) g(\dminus p \dplus q) e^{i q x}
	= \int \Haar{q} (f \cp g)(q) e^{i q x},
\end{align*}
where $\cp$ denotes the convolution product on $L^1(G_\ell)$. In other words, the only definition of the Fourier transform \eqref{eq:Md_Ft} suffices to write the ``quantizing'' relation
\begin{align}
	f \ g
	&= \mathcal{F}^{-1} \big( \mathcal{F}(f) \cp \mathcal{F}(g) \big).
	\label{eq:Md_spdef}
\end{align}

\paragraph{}
The Dirac delta function $\delta$ on the position space is defined as usual. We build a Dirac delta function, named also $\delta$ for simplicity, in the momentum space as the function that satisfies $\delta(p) = 0$ for any $p \neq 0$ and $\int \Haar{p} \delta(p) = 1$. It can equivalently be introduced as a measure on the topological group $(G_\ell, \dplus)$. Then, one computes that for $f \in L^1(G_\ell)$, one has
\begin{align*}
	\int \Haar{p} \delta(p) f(p)
	&= f(0)
	= \frac{1}{(2\pi)^{d+1}} \int \tdl{x} e^{i (\dminus 0) x} f(x)
	=  \frac{1}{(2\pi)^{d+1}} \int \Haar{p} \tdl{x} e^{ipx} f(x) \\
	&=  \frac{1}{(2\pi)^{d+1}} \int \Haar{p} \left( \int \tdl{x} e^{ipx} \right) f(p),
\end{align*}
which implies that
\begin{align}
	\delta(p)
	&=  \frac{1}{(2\pi)^{d+1}} \int \tdl{x} e^{i p x}.
\end{align}
The properties of this delta function have been derived in \cite{Hersent_2024a}. We summarise here the most important ones. Due to the noncommutativity of the addition law of momenta, one has that $\delta(p \dplus q) \neq \delta(q \dplus p)$, but thanks to a change of variable $p \to q \dplus p \dminus q$,\footnote{
	Note that we would have obtained the same result by integrating over $q$ and performing the change of variable $q \to \dminus p \dplus q \dplus p$.
}
one computes that
\begin{align*}
	\int \Haar{p} \delta(p \dplus q)
	&= \int \Haar{p} \mathscr{I}(q) \delta(q \dplus p)
\end{align*}
that we interpret directly as
\begin{align}
	\delta(p \dplus q)
	&= \mathscr{I}(q) \delta(q \dplus p).
	\label{eq:Md_dcyc}
\end{align}
In addition, one obtains from the definition of $\delta$ that
\begin{align*}
	\int \Haar{p} f(p) \delta(p \dplus q)
	&\overset{(p \to p \dminus q)}{=} \int \Haar{p} \mathscr{I}(q) f(p \dminus q) \delta(p)
	= \mathscr{I}(q) f(\dminus q),
\end{align*}
which implies the following properties
\begin{subequations}
\begin{align}
	\int \Haar{p} f(p) \, \delta(p \dplus q)
	&= \mathscr{I}(q) f(\dminus q), \\
	\int \Haar{p} f(p) \, \delta(q \dplus p)
	&= f(\dminus q).
\end{align}
\end{subequations}
In plain words, one should be careful with respect to which variable (left or right) the delta function is integrated over, the ``noncommutativity'' of the two expressions being captured by the modular function. By performing changes of variable, one can also obtain the following
\begin{align}
	\int \Haar{p} f(p) \delta(p \dminus q)
	&= \int \Haar{p} f(p) \delta(q \dminus p)
	= \mathscr{I}(\dminus q) f(q), \\
	\delta(\dminus p)
	&= \delta(p).
\end{align}

Finally, the cyclicity property of the delta function \eqref{eq:Md_dcyc} can be used to obtain a cyclicity property of the space integral. Indeed,
\begin{align}
\begin{aligned}
	\int \tdl{x} (fg)(x)
	&= \int \Haar{p} \Haar{q} \tdl{x} f(p) g(q) e^{i(p \dplus q)x} \\
	&= (2\pi)^{d+1} \int \Haar{p} \Haar{q} f(p) g(q)  \delta(p \dplus q) \\
	&= (2\pi)^{d+1} \int \Haar{p} \Haar{q} f(p) g(q) \mathscr{I}(q) \delta(q \dplus p) \\
	&= \int \tdl{x} (\mathscr{I}(g) f)(x)
\end{aligned}
	\label{eq:Md_cyc}
\end{align}
where $\mathscr{I} : \mathbb{A}_\ell \to \mathbb{A}_\ell$, in the last line, is the space operator that corresponds to the inverse Fourier transform of a multiplication by $\mathscr{I}$. 

The property \eqref{eq:Md_cyc} is sometimes referred to as twisted cyclicity of $\omega$ and $\mathscr{I}$ as the twist. Indeed, we show in appendix \ref{sapx:KMS_ro} that $\mathscr{I}$ is a regular automorphism of $\mathbb{A}_\ell$. The twist of cyclicity first appeared by studying field theory on $\kappa$-Minkowski and was considered a plague for the construction of gauge theory\footnote{
	For a review, see section 6 of \cite{Hersent_2023a}.
}.
This twist was used as a tool to build a differential calculus inducing a $\kappa$-Poincar\'{e} invariant gauge theory on $\kappa$-Minkowski \cite{Mathieu_2020} (also reviewed in \cite{Hersent_2023a}). The same reference also point out the similarities of $\mathscr{I}$ and the twist used in the context of twisted spectral triples, a matter we evoke in the concluding section \ref{subsec:c_op}.

\paragraph{}
Let us go back to the definition of a norm and the $C^*$-algebra axioms. By defining the inner product
\begin{align}
	(f, g)
	= \omega(f g^\dagger)
	&= \int \tdl{x} (f g^\dagger)(x),
	\label{eq:Md_ip}
\end{align}
for any $f, g \in \mathbb{A}_\ell$, one obtains that $\mathbb{A}_\ell$ is a Hilbert space and so a Banach space, for the associated norm
\begin{align}
	\Vert f \Vert
	= \omega(f f^\dagger)
	&= \sqrt{\int \tdl{x} (f f^\dagger)(x)}.
	\label{eq:Md_norm}
\end{align}
One of the non-trivial point to show is that this inner product \eqref{eq:Md_ip} is non-degenerate. It follows from the fact that $\omega$ is faithful, \ie $\omega(f f^\dagger) = 0$ if and only if $f = 0$. Indeed, one can write
\begin{align*}
	\omega(f f^\dagger)
	&= \int \tdl{x} \Haar{p} \Haar{q} f(p) \overline{f}(q) e^{ip x} e^{i (\dminus q) x}
	= (2\pi)^{d+1} \int \Haar{p} \Haar{q} f(p) \overline{f}(q) \delta(p \dminus q) \\
	&= (2\pi)^{d+1} \int \Haar{p} |f(p)|^2 \geqslant 0,
\end{align*}
which is $0$ if and only if $f(p) = 0$ for all $p \in G_\ell$ and so if and only if $f = 0$. This implies that we can define $\mathcal{H}_\omega = \mathbb{A}_\ell$ to be a Hilbert space on which $\mathbb{A}_\ell$ represent via the left multiplication, $\pi_\omega : \mathbb{A}_\ell \to \mathcal{L}(\mathcal{H}_\omega)$, with $\pi_\omega(f)g = fg$, for any $f, g \in \mathbb{A}_\ell$. Note that $\pi_\omega$ does not send elements of the algebra to bounded operators since even if $\Vert g \Vert < + \infty$, we may have $\Vert fg \Vert = +\infty$. However, one can still go on and define the operator norm on $\mathcal{L}(\mathcal{H}_\omega)$ as
\begin{align*}
	\Vert f \Vert_{\mathcal{L}(\mathcal{H}_\omega)}
	= \underset{\Vert g \Vert = 1}{\sup} \Vert fg \Vert.
\end{align*}
The property that $\mathbb{A}_\ell$ becomes a $C^*$-algebra when equipped with this norm is known, see for example section 2.1 of \cite{Landsman_1998} or appendix B.1.2 of \cite{Hersent_2024b} for a pedagogical introduction. Note that usually, the $C^*$-algebra built that way correspond to bounded operators $\mathcal{B}(\mathcal{H}_\omega)$, but we cannot here ensure boundedness of operators since $\omega$ is not normalised as mentioned in section \ref{subsec:tt_Md}.

\section{Thermal time from momentum space modular function}
\label{sec:tt}
\paragraph{}
In this section, we introduce the notion of thermal time together with its general construction. This is then applied to a generic deformed Minkowski spacetime, for which the structure of the momentum space is abundantly used. We also study the commutative limit of such a model, leading to an unthermalised Minkowski spacetime. Finally, some elements in the thermal time construction, that may be relevant for physical interpretation, are discussed.

\subsection{The thermal time hypothesis}
\label{subsec:tt_tt}
\paragraph{}
The notion of thermal time has been introduced in \cite{Rovelli_1993a, Rovelli_1993b} in order to solve the ``problem of time'' in a generally covariant quantum theory (see for example \cite{Anderson_2012} for a review). The latter problem actually arise as a conflict between the notion of time in quantum mechanics, or quantum field theory, and the notion of time in general relativity. In the former framework, the time evolution of states is performed thanks to a time flow that is common to all observers, whereas in the latter a time can only be defined locally and is thus observer dependant. With this theoretical clash comes the problem of defining a notion of time in a theory of quantum gravity. 

The thermal time hypothesis \cite{Connes_1994a} assumes that the time flow emerging from the generalised Heisenberg picture (corresponding to a $1$-parameter group of automorphism of a $C^*$-algebra in the following) of each state corresponds to the physical time flow of that precise state. The time is here interpreted as a pure parameter driving the dynamics (without any causal aspects) and is therefore called ``thermal'', due to its link with the way time is defined in statistical mechanics. Furthermore, this definition of time is \apriori state dependant but actually changes for different states in a controlled way.

\paragraph{}
As recalled in \cite{Connes_1994a}, the thermal time can be defined mathematically in the general framework of $C^*$-algebra thanks to major results of this theory: the Kubo-Martin-Schwinger (KMS) states, the Gel'fand-Na\u{\i}mark-Segal (GNS) construction, the Tomita-Takesaki theorem and the Connes' cocycle theorem. The full construction can be sketch as follows. We refer to textbooks like \cite{Connes_1994b, Landsman_1998, Gracia-Bondia_2001} for more detailed construction and definitions.

\paragraph{GNS construction.}
Given a $C^*$-algebra $\mathfrak{A}$, we consider a faithful state\footnote{
	Note that the GNS construction we are going to give here can also be done for state that are not faithful by quotienting $\mathfrak{A}$ by the ``kernel'' of the state, $J_{\psi_0} = \{ f \in \mathfrak{A}, \psi_0(f^\dagger f) = 0\}$.
}
$\psi_0$ (\ie a positive linear form of norm $1$) of $\mathfrak{A}$. Then, $\mathfrak{A}$ equipped with the inner product $\langle f, g \rangle_{\psi_0} = \psi_0(f^\dagger g)$ becomes a Hilbert space that we call $\mathcal{H}_{\psi_0}$. If one defines $\pi_{\psi_0} : \mathfrak{A} \to \mathcal{L}(\mathcal{H}_{\psi_0})$ to be the left multiplication, \ie $\pi_{\psi_0}(f)g = fg$, then one can show that it is a representation\footnote{
	Note that $\pi_{\psi_0}$ is faithful if and only if $\psi_0$ is faithful, and $\pi_{\psi_0}$ is irreducible if and only if $\psi_0$ is pure.
}
of $\mathfrak{A}$ on $\mathcal{B}(\mathcal{H}_\omega)$.

Physically, $\psi_0$ would be a vacuum state. We refer to \cite{Landsman_1998} on this point.

\paragraph{Tomita-Takesaki theorem.}
If one considers the conjugation ${}^\dagger$ of $\mathfrak{A}$ as a map from linear operators $\mathcal{L}(\mathcal{H}_{\psi_0})$ to itself through
\begin{align}
	\pi_{\psi_0}(f^\dagger)g
	&= \big(\pi_{\psi_0}(f)\big)^\dagger g
\end{align}
for any $f \in \mathfrak{A}$ and $g \in \mathcal{H}_{\psi_0}$, then ${}^\dagger$ admits a polar decomposition ${}^\dagger = J \triangle^{1/2}$, where $J$ is antiunitary and $\triangle$ is positive self-adjoint\footnote{
	Even if $\triangle$ is sometimes called the ``modular function'', it should not be confused \apriori with the modular function of the momentum space $\Delta$ defined earlier.
},
both as operators on $\mathcal{H}_{\psi_0}$. Given these elements, the Tomita-Takesaki theorem states that \begin{align}
	\alpha_t(\pi_{\psi_0}(f))
	= \triangle^{-it} \pi_{\psi_0}(f) \triangle^{it},
	\label{eq:tt_1pga}
\end{align}
for any $f \in \mathfrak{A}$ and any $t \in \Real$, forms a $1$-parameter group of automorphism of $\mathcal{B}(\mathcal{H}_{\psi_0})$. It is called the modular group of $\psi_0$. We have therefore built the thermal time associated with the state $\psi_0$.

\paragraph{Connes' cocycle theorem.}
However, the theory goes further, since the Connes' cocycle theorem states that $\{\alpha_t\}_{t \in \Real}$, defined in \eqref{eq:tt_1pga}, is actually independent of $\psi_0$, up to inner automorphsim. In other words, another state $\psi_1$ has a modular group given by $\{ u_t \alpha_t u_t^\dagger\}_{t \in \Real}$, for $u_t \in \mathcal{B}(\mathcal{H}_{\psi_0})$ unitary for any $t \in \Real$. The modular time is therefore defined for all states of $\mathfrak{A}$ up to this inner automorphism freedom. The impact of this freedom on the definition of the time and on the physics has not been explored in the literature and will be partially discussed in section \ref{subsec:tt_dis}.

\paragraph{KMS states.}
Finally, from another point of view, the $C^*$-algebra formalism can be used to the study of (quantum) statistical systems \cite{Connes_1994b}. Here, $\mathfrak{A}$ is the algebra of observables and we consider that there exists a $1$-parameter group of automorphism $\{\gamma_t\}_{t \in \Real}$ that controls the time evolution of the system. Usually, one writes $\gamma_t(f) = e^{itH} f e^{-itH}$, with $H$ the Hamiltonian of the system. In this context, an equilibrium state correspond to a so-called KMS state of $\mathfrak{A}$ with (inverse) temperature $\beta > 0$ \cite{Haag_1967}. Such a state $\omega : \mathfrak{A} \to \Cpx$ is defined, for any $f, g \in \mathfrak{A}$, by the existence of a function $F: \Cpx \to \Cpx$ such that it is holomorphic on the strip $\{z \in \Cpx, 0 \leqslant \Im(z) \leqslant \beta\}$ and that satisfies
\begin{align}
	F(t)
	&= \omega( f \gamma_t(g) ), &
	F(t + i\beta)
	&= \omega( \gamma_t(g) f )
	\label{eq:tt_KMS_def}
\end{align}
for any $t \in \Real$.

A very astonishing result (see for example \cite{Haag_1996} equation (V.2.14)) is that $\psi_0$ (defined above) satisfies the KMS condition \eqref{eq:tt_KMS_def} with the $1$-parameter group $\{\alpha_t\}_{t \in \Real}$ and temperature $\beta = 1$. Note that, in a non-relativistic regime, one can even show \cite{Connes_1994a, Haag_1996} that
\begin{align}
	\alpha_t = \gamma_{\beta t},
	\label{eq:tt_nr_temp}
\end{align}
so that the two notions of time (KMS and modular) are actually proportional, with proportionality constant being the (inverse) temperature $\beta$. This genuine observation actually relates the notion of (KMS) thermal time of a state (the parameter that drives its dynamics) to the notion of ``modular time'' (an algebraic notion \apriori) that happens to be state-independent. Being able to parametrize the dynamic in a state independent way is precisely what drove the authors of \cite{Connes_1994a} to postulate that this notion of thermal time is the physical time. 

\paragraph{}
Therefore, given any $C^*$-algebra, one can define a global notion of time, the thermal time, by following the previous procedure. This has led some authors to state that operator algebras are canonically dynamical (see for example the introduction of \cite{Connes_1994a}).

\subsection{Thermal time of deformed Minkowski spacetime}
\label{subsec:tt_Md}
\paragraph{}
In this section, we want to show that in the context of a general (noncommutative) deformation of Minkowski, as introduced in section \ref{sec:Md}, one can construct the thermal time and that it emerges from the momentum space modular function. In this section, we make the hypothesis that the momentum group $G_\ell$ is not unimodular, that is $\mathscr{I} \neq 1$, \ie it is not identically $1$ for any $p \in G_\ell$. Note that the unimodular case $\mathscr{I} = 1$ is discussed in section \ref{subsec:tt_dis}.

\paragraph{}
We consider here that our $C^*$-algebra is the deformed Minkowski space, \ie $\mathfrak{A} = \mathbb{A}_\ell$. In this sense, we are considering that our (quantum) system \emph{is} the spacetime itself and not some observer evolving on this spacetime.

We further consider $\psi_0 = \omega$. Note that $\omega$ is a weight and not a state because it cannot be normalised. Explicitly, $\omega(1) = \int \td x = \mathrm{Vol}(\Mink{d}) = +\infty$. However, one can show that considering a weight instead of a state is not changing the mathematical construction of section \ref{subsec:tt_tt}. One should pay attention though that ``expectation values'', of the form $\Vert f\Vert$, cannot be regarded as probabilities since they cannot be normalised.

\paragraph{GNS construction.}
We have already built $\mathcal{H}_\omega = \mathbb{A}_\ell$, with inner product $(\cdot, \cdot)$ defined in \eqref{eq:Md_ip}, and its representation in section \ref{subsec:Md_Cs}, by following the GNS construction detailed in section \ref{subsec:tt_tt}.

\paragraph{Tomita-Takesaki theorem.}
By going in momentum space, one can write
\begin{align*}
	f^\dagger(x) 
	&= \int \Haar{p} \overline{f}(p) e^{i(\dminus p)x}
	\overset{(p \to \dminus p)}{=} \int \Haar{p} \mathscr{I}(p) \overline{f}(p) e^{i p x}
	= \big(J \circ \mathscr{I}\big)(f)(x),
\end{align*}
where we noted $J$ the inverse Fourier transform of the complex conjugation in momentum space. One can show that $J$ is antiunitary for the inner product \eqref{eq:Md_ip}, \ie $(J(f), J(g)) = \overline{(f,g)} = (g,f)$, and that $\mathscr{I}$ is self-adjoint for this inner product, \ie $(\mathscr{I}(f), g) = (f, \mathscr{I}(g))$, for any $f, g \in \mathbb{A}_\ell$, by again going into momentum space. Therefore, the momentum space operators directly provide us with a decomposition ${}^\dagger = J \triangle^{1/2}$, with $\triangle = \mathscr{I}^2$. Note the troubling (but maybe mathematically irrelevant, according to section \ref{subsec:tt_dis}), presence of a factor $2$, that also induce a factor $2$ between our notion of thermal time and the one defined in \cite{Connes_1994a}.

\paragraph{Connes' cocycle theorem.}
In the present case, one can fully characterise the inner automorphisms. Indeed, by writing explicitly the unitarity condition $(u(f), u(g)) = (f,g)$ and going into momentum space, one obtains $u(p) \overline{u(p)} = 1$, \ie $u(p) \in U(1)$. This implies that all unitary elements are of the form $u(p) = e^{i F(p)}$ in momentum space, for an arbitrary $F: G_\ell \to \Real$. By inverse Fourier transform, one can write $u = e^{i F}$ where $F : \mathbb{A}_\ell \to \mathbb{A}_\ell$ is the position operator associated to $F(p)$. Further discussion on this point is postponed to section \ref{subsec:tt_dis}.

\paragraph{KMS state.}
We prove in the appendix \ref{apx:KMS} that $\omega$ is a KMS weight for the $1$-parameter group of automorphism $\{\mathscr{I}^{it}\}_{t\in\Real}$. Therefore, the thermal time of the deformed Minkowski spacetime can be defined thanks to the modular function of its momentum space. In the generalised Heisenberg picture, the associated Hamiltonian to this evolution is $\ln(\mathscr{I})$.

\paragraph{}
At this point, we wish to discuss the physical importance of this thermal time. Being a possible solution to the ``problem of time'' is of course the main asset of this theory in the context of quantum gravity (as discussed above), but may not be of primordial interest here. Indeed, we are dealing with deformations of Minkowski spacetime, which have a global time coordinate. Therefore, there is no problem of time here.

This setting allows us to interpret the noncommutativity of the spacetime as being thermal. In other words, the simple fact that the spacetime exists with noncommuting coordinates is sufficient\footnote{
	To be very precise, non-unimodularity of the momentum space is needed, otherwise the spacetime vacuum state is stationary. See section \ref{subsec:tt_cl} on this point.
}
to make the vacuum state $\omega$ of the spacetime evolve through time. The noncommutativity, generally interpreted as generating minimal observable surfaces (see section \ref{subsec:Md_Pd}) and so as ``fuzziness'', is here generating a thermal behaviour of the spacetime. If one is to interpret deformations of Minkowski as a ``flat limit'' of quantum gravity, then the quantum behaviour of gravity in this limit could be interpreted as been purely thermal. Put differently, in a similar way as the thermal treatment of a system in a bath models the many degrees of freedom constituting the bath, the quantum gravity degrees of freedom in its ``flat limit'' consists of this thermal behaviour of the spacetime.

Therefore, one could bridge the treatment of ``flat'' quantum gravity with the one of thermal field theory, from which many techniques and phenomenological considerations are available. We refer to the concluding section \ref{subsec:c_ph} for a more detailed discussion on phenomenology. Note that a correspondence between the thermal behaviour of the tracial weight has already been associated to gravitational degrees of freedom in the context of AdS/CFT \cite{Verlinde_2020}, or to an emergent spacetime in Ishibashi-Kawai-Kitazawa-Tsuchiya (IKKT) matrix models \cite{Brahma_2022}. Furthermore, the idea of defining a theory of quantum gravity as a generally covariant thermal field theory, where thermality encodes microscopic degrees of freedom has been explored in \cite{Kotecha_2019} (see references therein for thermal aspects of group field theory). Other quantum gravity approaches sparks a thermal behaviour like $2+1$-dimensional quantum gravity \cite{Smolin_1995} or spacetime foam \cite{Garay_1999}. As a side remark, it can be noted that the introduction of a temperature can be used as a computational technique, known as ``hard thermal loop expansion'', for the quantisation of gravity, see for example \cite{Brandt_1998, Park_2021, Brandt_2023}.

\subsection{Commutative limit of the thermal time}
\label{subsec:tt_cl}
\paragraph{}
The commutative limit corresponds, in the physics literature, to the formal limit when the deformation parameter vanishes, here $\ell \to 0$, and thus tends to the undeformed theory on Minkowski spacetime with Poincar\'{e} symmetries. The rigorous  mathematical commutative limit may be more complex then that, but it goes far beyond the scope of this paper. Here, we consider that all functions on the deformed spaces can be developed formally as infinite power expansions in $\ell$ and we define the ``commutative limit'' as the limit for which only the zeroth order of this expansion remains.

\paragraph{}
In this limit, one has that the algebra of functions on Minkowski becomes the commutative algebra of its smooth functions $\mathbb{A}_\ell \to C^\infty(\Mink{d})$ and the vector space spanned by its global coordinates simply Minkowski $\Mink{d}_\ell \to \Mink{d}$. The weight $\omega$ becomes the spacetime integral on Minkowski. In a similar way, the vector space spanned by the global momentum coordinates is also Minkowski $G_\ell \to \Mink{d}$ and its additive composition law can easily be shown, from its definition of section \ref{subsec:Md_pm}, to be the usual addition $\dplus = +$. It follows that the Haar measure on $G_\ell$ reduces to the Lebesgue measure $\td p$ with unimodularity property $\mathscr{I} = 1$.

A direct consequence of the unimodularity is that there are no thermal evolution of $\omega$, and so of Minkowski spacetime. In other words, in the commutative limit, one has $\alpha_t = \id$, for any $t \in \Real$. The fact that commutative orientable $4$-manifolds have a trivial modular group for the global trace has already been noticed in \cite{Etesi_2024}. Note that this is also consistent with the early work \cite{Rovelli_1993b} where the thermal time of the Friedmann-Lema\^{i}tre-Robertson-Walker universe coupled to an electromagnetic field is derived. The dynamics of such a classical spacetime is non-trivial precisely thanks to the electromagnetic field: from equation (25), the temperature is $0$ if the electromagnetic Hamiltonian is changed to $0$. Finally, we wish to point out that the trivial evolution of a classical spacetime is coherent with the fact that the spacetime (here Minkowski) is not supposed to be ``quantum'' anymore, and so to be free of thermal source\footnote{
	Let us underline once more that we are characterising here the thermal behaviour \emph{of} the spacetime itself. Even on classical geometries can observers \emph{on} the spacetime experience a temperature like in the Unruh effect.
}.

\subsection{Discussion on the thermal time construction}
\label{subsec:tt_dis}
\paragraph{}
We want to discuss in the following paragraphs two matters that have not been discussed in the literature, as far as we know.

\paragraph{}
First, one should note that nothing prevents that $\triangle = 1$. In this case, $\alpha_t = \id$ for any time $t$ and there are ``no'' dynamics in the sense that $\psi_0$ will remain unchanged throughout the (thermal) time evolution. In a non-relativistic picture, it amounts to have a trivial Hamiltonian associated to this modular group, \ie $H = \ln(\triangle) = 0$. In this precise case, the vacuum state $\psi_0$ cannot be used to define a time flow. This remark is of importance since it concerns all classical spacetimes \cite{Etesi_2024} as detailed in section \ref{subsec:tt_cl}, as well as unimodular deformation of Minkowski, discussed in section \ref{subsec:kM_uni}.

\paragraph{}
Furthermore, the uniqueness of the modular group is valid up to inner automorphisms, as discussed earlier. However, there has not been, as far as the author knows, any discussion on ``how much'' this inner automorphism freedom could change the physics of different states.

A first and simple statement is to notice that for a given $\lambda \in \Real$, $\triangle^{i \lambda}$ defines a unitary element. Therefore, one could consider to generate an inner automorphism with $u_t = \triangle^{-i \beta(t)}$ for any continuous function $\beta : \Real \to \Real$. In this respect, while the vacuum state $\psi_0$ would experience a time evolution given by $\alpha_t = \triangle^{-it} \cdot \triangle^{it}$, another state could experience a time evolution given by $\tilde{\alpha}_t = u_t \alpha_t u_t^\dagger = \triangle^{-i\tilde{\beta}(t)} \cdot \triangle^{i\tilde{\beta}(t)}$. Therefore, the inner automorphism freedom can induce a complete redefinition of the thermal time (and possibly of the temperature) between two different states.

In the case of Minkowski deformations, the set of inner automorphisms has been fully characterised to be generated by unitary elements of the form $u(p) = e^{i \beta(t, p)}$, where $\beta: \Real \times G_\ell \to \Real$ is an arbitrary function in momentum space. By writing the time evolution of $\omega$ in a generalized Heisenberg picture as $\alpha_t = e^{-i t H} \cdot e^{i t H}$, where $H = \ln(\mathscr{I})$ is the generalized Hamiltonian, the inner automorphism transformation writes
\begin{align}
	\tilde{\alpha}_t
	&= e^{-i (t H + \beta(t)) } \cdot e^{i(t H + \beta(t))},
	\label{eq:tt_Md_iai}
\end{align}
with $\beta(t): \mathbb{A}_\ell \to \mathbb{A}_\ell$. In this case, the new state can experience another (arbitrary) reparametrisation of time and/or of the Hamiltonian for its dynamics. This observation seems to be of primordial importance for the interpretation of the physics of the modular theory but is very puzzling since it goes against the usual treatment of the physics of quantum thermodynamics: the Hamiltonian is given as an entry data for the full system, whatever the states of its constituents. See for example \cite{Potts_2019} for an introduction.

\paragraph{}
From this short analysis, it seems that the inner automorphism transformation could have a dramatic impact on the physical systems under study. Yet, a consistent approach to the study of physical consequences due to these transformations would consist of considering explicit examples of states first. Indeed, the unitary element $u_t$ is fixed as soon as the new state $\psi_1$ has been chosen. Therefore, the reparametrisation of time and Hamiltonian may make sense physically from the point of view of this new state. The characterisation of the inner automorphism group and its physical consequences are left open for future dedicated work.

\section{Thermal behaviour of \tops{$\kappa$}{kappa}-Minkowski}
\label{sec:kM}

\subsection{Thermal time construction}
\paragraph{}
The $\kappa$-Minkowski spacetime corresponds to a so-called $\kappa$-deformation of Minkowski that has been established during the 90's. It was first built via the bicrossproduct construction \cite{Majid_1994} of the $\kappa$-Poincar\'{e} quantum group built in \cite{Lukierski_1991}. Its (Lie) algebra of coordinates $\Mink{d}_\kappa$ satisfies
\begin{align}
	[x^0, x^j]
	&= \frac{i}{\kappa} x^j, &
	[x^j, x^k]
	&= 0,
	\label{eq:kM_coord}
\end{align}
where $j,k = 1, \ldots, d$ are space indices and $\kappa > 0$ is the deformation parameter, that has mass unit. Here, $x^\mu$ denotes the (global) coordinate function that are dual to the translation generators $P_\mu$ of $\kappa$-Poincar\'{e}. A more detailed definition and historical construction of $\kappa$-Minkowski can be found in \cite{Hersent_2023a} (see also the references therein). 

The Lie algebra \eqref{eq:kM_coord} is called the affine Lie algebra and has, as associated Lie group, the affine group $G_\kappa = \Real \ltimes \Real^d$, corresponding here to the momentum space of $\Mink{d}_\kappa$. It can be shown (see for example \cite{Williams_2007}) that this group is not unimodular and the modular function is given by $\mathscr{I}(p) = e^{- d p_0/\kappa}$, where $d$ is the space dimension of the $d+1$-dimensional spacetime. One can perform the analysis of sections \ref{sec:Md} and \ref{sec:tt} to show that $\omega$ (the spacetime integral) is a faithful weight on $\kappa$-Minkowski $\mathbb{A}_\kappa$, together with the fact that it is a KMS weight for the $1$-parameter group of automorphism $\{\mathscr{I}^{it}\}_{t\in\Real}$. Note that $\omega$ was first introduced as a KMS weight with modular operator $\mathscr{I}$ in \cite{Matassa_2013, Poulain_2018}. Moreover, one can find an explicit characterisation of the $C^*$-algebra structure of $\mathbb{A}_\kappa$ in \cite{Durhuus_2013, Poulain_2018}. Note that the last two references show that $\omega$ is $\kappa$-Poincar\'{e} invariant, \ie $\omega( M \actl f) = \varepsilon(M) \omega(f)$, for any $M \in \Poin{d}_\kappa$ and $f \in \mathbb{A}_\kappa$, where $\varepsilon$ denotes the counit of $\Poin{d}_\kappa$.

\paragraph{}
By explicit computations using the Fourier transform, it can be shown \cite{Poulain_2018} that the position operator associated to $\mathscr{I}$ act on functions as $\mathscr{I}(f)(x^0, x^j) = f(x^0 + d\frac{i}{\kappa}, x^j)$. Therefore, the time evolution operator $\mathscr{I}^{-it}$ acts, for any $t \in \Real$, as
\begin{align}
	\mathscr{I}^{-it}(f)(x)
	&= f( x^0 + \frac{d}{\kappa} t, x^j).
	\label{eq:kM_tt}
\end{align} 
It therefore corresponds to a translation with respect to the global time $x^0$ of a time $\frac{d}{\kappa} t$. In the context of the $\kappa$-Minkowski space, we have thus shown that the thermal time evolution corresponds to a translation in the global time coordinate, in full accordance with the thermal time hypothesis of \cite{Connes_1994a}. Several comments are of importance here.

\begin{enumerate}[label=(\roman*)] 
	\item Having characterised the thermal time, one can associate its factor, here $d/\kappa$ to an (inverse) temperature. It is even more tempting to define the temperature as being $\kappa$ and let $d$ be a geometrical prefactor\footnote{
		In \cite{Rovelli_1993b}, the time flow has an unexpected prefactor $\frac{3}{4}$. Linking this to our analysis, it may be that the factor $3$ corresponds to $d$, the spacial dimension of the spacetime. The $1/4$ factor may be due to the author's introduction of a factor $4$ in equation (24). We agree that such a geometrical interpretation needs to be made more explicit.
	}.
	The relation between $\kappa$ and the temperature has already been found in a related context \cite{Kowalski-Glikman_2009}, but without the notion that $\kappa$ actually corresponds to the global temperature of the spacetime.
	\item If one is to interpret this temperature as coming from quantum gravity effect, then $\kappa$ may be related to the Planck scale. The temperature of $\omega$ would thus be the Planck temperature (around $1.4 \times 10^{32}$ K), which makes $\omega$ an ``infinite temperature state'' from the classical point of view (\ie in the commutative limit $\kappa \to + \infty$). The fact that the tracial\footnote{
		From a purely mathematical point of view, $\omega$ is not a trace because it is not cyclic: $\omega(fg) \neq \omega(gf)$. However, it is a \emph{twisted} trace as it satisfies the \emph{twisted} cyclicity \eqref{eq:Md_cyc}, where $\mathscr{I}$ is called the ``twist''. We want to underline, as a side remark, that the fact that the twist is non-trivial, \ie $\mathscr{I} \neq 1$, is precisely what makes the thermal dynamics non-trivial, \ie $\alpha_t \neq \id$.	
	}
	state (or weight) corresponds to the highest temperature state was actually pointed out in \cite{Etesi_2024}. We are therefore considering a high temperature state to define the thermal time of all other states of the spacetime.
\end{enumerate}

As a side remark, we wish to put forward that our analysis is one more step in reconciling the quantum group approach and the $C^*$-algebra approach, contrary to the criticism of \cite{Majid_2005}.

\subsection{\tops{$\kappa$}{kappa}-Poincar\'{e} invariance}
\paragraph{}
The thermal time $t$ defined above is \emph{not} $\kappa$-Poincar\'{e} invariant, in the sense that another inertial frame (related to the one of $x^\mu$ by a $\kappa$-Poincar\'{e} transformation) does not experience a time flow given by $t$, as viewed from the initial frame. In other words, the thermal flow is a global time translation and not a proper time translation.

Explicitly, let us call $x^{\mu\prime} = \tensor{\Lambda}{^\mu_\nu} \otimes x^\nu + a^\mu \otimes 1$ the coordinates of the new frame after a $\kappa$-Poincar\'{e} transformation (see \cite{Mercati_2023} for the algebra relations of $\Lambda$ and $a$), then one can show that the $x^{\mu\prime}$'s also satisfies \eqref{eq:kM_coord}. The associated momentum space coordinates are defined as above and read $p_\mu^\prime = \tensor{(\Lambda^{-1})}{_\mu^\nu} \otimes p_\nu$, with a modular function $\mathscr{I}(p^\prime) = e^{-d p_0^\prime / \kappa}$. It implies that the thermal time flow follows \eqref{eq:kM_tt} in those new coordinates, \ie
\begin{align*}
	\mathscr{I}^{-it}(f)(x^\prime)
	&= f(x^{0\prime} + \frac{d}{\kappa} t, x^{j\prime})
	= f\left( \tensor{\Lambda}{^0_\mu} \otimes x^\mu + 1 \otimes \frac{d}{\kappa} t + a^0 \otimes 1, \tensor{\Lambda}{^j_\mu} x^\mu + a^j \otimes 1 \right).
\end{align*}
Therefore the thermal time flow corresponds to a shift of $\beta t$ of the time coordinate $x^{0\prime}$ in the new frame. The reason of this variance simply follows from the fact that the thermal Hamiltonian considered here $- \ln(\mathscr{I}) = \frac{d}{\kappa} p_0$ is not $\kappa$-Poincar\'{e} invariant. Equivalently, one can check that $\mathscr{I}$, viewed as an element of $\Poin{d}_\kappa$ is not central,
\begin{align*}
	[\mathscr{I}, P_\mu] = 0, &&
	[\mathscr{I}, J_j] = 0, &&
	[\mathscr{I}, K_j] = i \frac{d}{\kappa} P_j \mathscr{I} \neq 0,
\end{align*}
and that the transformations impacting the thermal time corresponds to the boosts\footnote{
	To be fully complete, even at the level of the classical theory, the (global) time translation, generated by $P_0$, are not Poincar\'{e} invariant since $[P_0, K_j] = i P_j \neq 0$. Therefore, the notion of time evolution commonly used in quantum field theory is $\alpha_t = e^{-i t H} \cdot e^{i t H}$ where $H$ is a Poincar\'{e} invariant Hamiltonian, like $H(p) = \sqrt{p^2 + m^2}$ for a single massive particle. This allows for a time flow independent of the inertial frame.
}.

\paragraph{}
The previous analysis shows that the thermal time evolution only corresponds to the (global) time evolution in a preferred frame. This a common feature of thermal field theory (see for example section 2 of \cite{Salvio_2025}), where the preferred frame can be shown to be the rest frame of the system.

Note that this $\kappa$-Poincar\'{e} breaking is not hampering us from performing field theory predictions when interpreting the noncommutativity of the spacetime as a thermal source. Indeed, as one can show (see for example section 3.3 of \cite{Weinberg_1995}) that a non-covariant Hamiltonian does not prevent one from having a covariant S-matrix.

\subsection{Unimodular momentum space and discrete time}
\label{subsec:kM_uni}
\paragraph{}
There exists unimodular (noncommutative) deformations of Minkowski. As major examples, one can cite the Moyal space or $\rho$-Minkowski. The explicit computation of their momentum group and their properties can be found for example in \cite{Hersent_2023a} and \cite{Hersent_2023c} respectively. Therefore, for those spacetime, the tracial weight $\omega$ has a trivial thermal evolution $\alpha_t = \id$. It implies that one cannot define a thermal time by considering only the tracial weight. The case of $\rho$-Minkowski is quite interesting thanks to its close relation with the case of $\kappa$-Minkowski and allows us to relate the trivial thermal time evolution with the discreteness of time.

\paragraph{}
The fact that $\rho$-Minkowski has a discrete time (spectrum) has first been noticed in \cite{Lizzi_2021} and can be understood quite simply. Compared to $\kappa$-Minkowski having the affine group $\Real \ltimes \Real^d$ as momentum space, the $\rho$-Minkowski space is built out of the euclidean group, that one can write $U(1) \ltimes \Real^d$. One can actually go from the $\kappa$-deformation to the $\rho$-deformation by performing a compactification $\rho \leftrightarrow i/\kappa$ \cite{Hersent_2023c}. In terms of group elements, seen in a matrix representation, one has
\begin{align}
	W_\kappa(p_0, p_j) &=
	\begin{pmatrix}
		e^{- p_0 / \kappa } & 0 \\
		0 & p_j
	\end{pmatrix} &
	\longleftrightarrow &&
	W_\rho(p_0, p_j) &=
	\begin{pmatrix}
		e^{i \rho p_0} & 0 \\
		0 & p_j
	\end{pmatrix},
\end{align}
where the compactification occurs as the real exponential becomes a complex exponential, that is $\Real$ is mapped to $U(1)$. At the level of the spacetime, this is directly visible since one go from $\kappa$ to $\rho$ by considering $x^j \leftrightarrow e^{i \varphi}$, where $\varphi$ is an angular variable. As a direct consequence of the compactification, the associated ``momenta'' to $\varphi$ are integers of the Fourier series expansion. The fact that $x^0$ and $\varphi$ do not commute in $\rho$-Minkowski therefore impose that they cannot be co-diagonalised so that if one uses the (discrete) basis of eigenstates of $\varphi$, then $x^0$ ends up having a discrete spectrum. Note that from this point of view, $\rho$ corresponds to the time step of this discrete time.

The relation between the compactification and the fact that the modular function becomes trivial is also quite simple. For $\kappa$-Minkowski, the modular function reads $\mathscr{I}_\kappa(p) = e^{-d p_0 / \kappa}$ which becomes, after compactification\footnote{
	A doubtful reader can consult \cite{Hersent_2023c} for the computation of $\mathscr{I}_\rho$ without the use of $\mathscr{I}_\kappa$.
},
$\mathscr{I}_\rho(p) = \vert e^{i d \rho p_0} \vert = 1$. This is also fully compatible with the fact that, since the time needs to be discrete, there cannot be a \emph{continuous} $1$-parameter group of automorphism parametrizing the time flow.

\paragraph{}
Beyond the full characterisation of unimodular cases, the example of $\rho$-Minkowski allows us to disentangle the unimodularity coming from the fact that the spacetime is classical (in which case the time is continuous) and the unimodularity due to a new property of the time in such noncommutative spacetime. Moreover, the absence of ``thermality'' in unimodular quantum spacetimes can still have phenomenological consequences, as discussed in \cite{Lizzi_2021} concerning the discreteness of time.

\section{Conclusion and discussion}
\label{sec:c}

\subsection{Summary of the results}
\label{subsec:c_sum}
\paragraph{}
We summarise here the main results of the paper.
\begin{enumerate}[label=(\roman*)]
	\item The thermal time of \cite{Connes_1994a} has been constructed for any noncommutative (quantum) deformation of the Minkowsi spacetime. It is defined as being the modular group flow of the (high temprature) tracial weight, where the modular operator corresponds to the inverse Fourier transform of the momentum space modular function.
	\item The commutative limit of the modular group corresponds to the identity such that we are left with an unthermalised (classical) Minkowski spacetime.
	\item The thermal time flow is derived in the case of $\kappa$-Minkowski and correspond to a global time translation. The global time not being $\kappa$-Poincar\'{e} invariant, it is showed that the thermal time is not either.
	\item The unimodularity of $\rho$-Minkowski, leading to an undefined thermal time for the tracial weight, is shown to be directly related to the discreteness of time in this model.
\end{enumerate}
This work consists of a first step in the thermal properties of noncommutative spacetimes and its physical consequences. Therefore, it opens the way for many new considerations, but also brings a lot of unsolved problems as discussed in the following sections.

\subsection{Phenomenological consequences}
\label{subsec:c_ph}
\paragraph{}
The study of noncommutative spacetimes has already sparked a tremendous amount of quantum gravity phenomenological consideration in the scope of highly energetic cosmic messengers. We refer to the review \cite{Addazi_2022} for more details. In the search for other quantum gravity phenomenology, recent hopes are also focusing on tabletop experiments (see for example section 2.10 of \cite{Buoninfante_2024}). However, as far as the author is aware of, there has been very few works in thermal field phenomenology for quantum gravity. We here summarise some of the aspects of thermal field phenomenology that may be relevant in the scope of quantum gravity.

\paragraph{}
If one considered that a particle is coupled to a thermal bath, then it could happen that this particle is produced out (or absorbed by) the bath itself (see \cite{Salvio_2025} section 5). In our line of thought, the ``bath'' consists only of the (noncommutative) spacetime and the coupling would simply be that these particles live on this spacetime (\ie gravitation). Therefore, from the simple fact that the spacetime is noncommutative, one could expect that particles are produced/absorbed by it. This phenomenon has already been computed for a gauge theory on $\kappa$-Minkowski, for which the one-point function (\ie a photon being emitted/absorbed from nothing) at one loop is shown to be non-trivial \cite{Hersent_2022a}. At the level of the (quantum) field theory, this property is quite problematic since it implies that the vacuum of the theory is unstable under quantum fluctuations. From the thermal point of view thought, a non-zero one-point function makes perfect physical sense.

From these considerations, it could be that particles (like protons or photons) have a ``lifetime'' corresponding to their mean time of flight before absorption by the bath. It is important to note that the actual bounds for the proton lifetime (of around $9\times 10^{29}$ years \cite{PDG_2024}) cannot be applied in this case, since it does not consists of a decay.

A complementary view on this point is the possible effect of the thermal bath on the particle masses. The self-energy of these particles may be affected by the bath and therefore change their (effective) mass. One of the most discussed topic on this point concerns the possibility of a photon mass. In a model based on a deformation of Minkowski, it could be that a massless photon obtains an effective mass due to its coupling with the bath. No such mass has been derived in noncommutative gauge theories because of the UV/IR mixing effect (see for example \cite{Hersent_2023a, Hersent_2024a}.

\paragraph{}
From another perspective, thermal field theories may experience phase transitions (see section 6 of \cite{Salvio_2025}), and thus display different behaviours for different temperatures. In our ``flat limit'' model, the temperature of the vacuum state is presumably constant, as in the case of $\kappa$-Minkowski, which has a temperature $\kappa$ throughout all spacetime. However, in a more advanced (or maybe curved) version of such models, if a thermal picture can also be built, the temperature may become spacetime dependent and therefore generate phase transitions.

\paragraph{}
Also quite promising, these properties of a noncommutative spacetime needs to be derived properly before any analysis could be performed. This is a task we leave for future work.

\subsection{Open problems}
\label{subsec:c_op}
\paragraph{}
Finally, we resume here the problems that have been left open in this article, not counting the ones of the previous section.

\paragraph{}
First, the notion of (thermal) time, constructed in section \ref{sec:tt}, is independent of the state under consideration, up to an inner automorphism transformation of the $1$-parameter group of automorphism parametrising the time flow of the states. Meaning that two different states have the same time flow morphism up to the action of a unitary element. The physical implication, either for the thermal time or for the Hamiltonian of the system, of such a transformation has been poorly studied. A brief argument developed in section \ref{subsec:tt_dis} showed that this transformation could have a tremendous impact on the physical relevance of the considered quantities. This needs however to be deepened. 

\paragraph{}
As analysed in this paper,  the non-unimodularity condition is essential for defining the thermal time. However, some noncommutative spacetime are unimodular and therefore without thermal time. It has been shown in the case of the (unimodular) $\rho$-Minkowski space that the absence of thermal time can be directly linked to the fact that its global time coordinate is discrete. The differences between unimodular and non-unimodular deformations has already been experienced in noncommutative field theory, like the appearance of a twist in the cyclicity of the trace\footnote{
	From \eqref{eq:Md_cyc}, the integral is cyclic if and only if $\mathscr{I} = 1$, that is if and only if the spacetime is unimodular.
}
which has caused trouble for the definition of a gauge theory on $\kappa$-Minkowski (see for example section 6 of \cite{Hersent_2023a}), or the fact that unimodular spacetimes simplify considerably the definition of quantum field theories \cite{Fabiano_2025}. However, there is no fundamental comprehension of these discrepancies yet.

\paragraph{}
It has been already noticed \cite{Poulain_2018} that the modular function of a deformed Minkowski space fulfils all the requirements of the twist in the context of twisted spectral triples. It is yet unclear if the twist has some relevance in the study of the momentum space of those theories, maybe as a generalised modular function. Moreover, it has not been studied whether the twist can be used to define a $1$-parameter group of automorphism on the spectral triple, and therefore to define a thermal time. Such consideration may be relevant since the (twisted) spectral triples consists, as such, of generalised Riemannian (and therefore timeless) geometries.

\section*{Acknowledgement}
The author thanks J.~C.~Wallet for early discussions on this work. Comments from F.~Mercati and J.~de Ramon Rivera were greatly appreciated. We also thank C.~Rovelli for reading suggestions.

\noindent%
This work has been supported by the grants CNS2023-143760, funded by NextGenerationEU and PID2023-148373NB-I00 funded by MCIN/AEI/10.13039/501100011033/FEDER – UE.

\noindent%
This work falls within the scopes of the COST Action CA23130 ``Bridging high and low energies in search of quantum gravity'' and the COST Action 21109 CaLISTA ``Cartan geometry, Lie, Integrable Systems, quantum group Theories for Applications'', from the European Cooperation in Science and Technology (COST).

\noindent%
\textbf{This is a preliminary version. Any comments or questions are welcome.}

\appendix
\section{The spacetime integral as a KMS weight}
\label{apx:KMS}

\subsection{Proof that \tops{$\mathscr{I}$}{I} is a regular automorphism of \tops{$\mathbb{A}_\ell$}{Al}}
\label{sapx:KMS_ro}
\paragraph{}
We show here that $\mathscr{I} : \mathbb{A}_\ell \to \mathbb{A}_\ell$, defined in section \ref{subsec:Md_Cs}, is a regular automorphism. Explicitly, $\mathscr{I}$ is an automorphism $\mathscr{I}(fg) = \mathscr{I}(f) \mathscr{I}(g)$, that is regular $\mathscr{I}(f^\dagger) = (\mathscr{I}^{-1}(f))^\dagger$, for any $f,g \in \mathbb{A}_\ell$. Both are proven by going to momentum space and using the properties of the modular function.

First, one has
\begin{align*}
	\big(\mathscr{I}(f) \mathscr{I}(g)\big)(x)
	&= \int \Haar{p} \Haar{q} \mathscr{I}(p) \mathscr{I}(q) f(p) g(q) e^{ipx} e^{iqx} \\
	&= \int \Haar{p} \Haar{q} \mathscr{I}(p \dplus q) f(p) g(q) e^{i(p \dplus q) x} \\
	&\!\!\!\!\!\!\!\! \overset{(q \to \dminus p \dplus q)}{=} \int \Haar{p} \Haar{q} \mathscr{I}(q) f(p) g(\dminus p \dplus q) e^{i q x} \\
	&= \int \Haar{q} \mathscr{I}(q) (f \cp g)(q) e^{i q x} \\
	&= \mathscr{I}(fg)(x),
\end{align*}
where the last equality holds thanks to \eqref{eq:Md_spdef}. Note that we have also used the fact that the modular function is a morphism, that is $\mathscr{I}(p) \mathscr{I}(q) = \mathscr{I}(p \dplus q)$, for any $p, q \in G_\ell$.

Concerning the second relation, one need first to prove that the $\dminus$ law in momentum space is related to the involution ${}^\dagger$ of the position space. This is done through the general relation of module $*$-algebra theory
\begin{align}
	(P_\mu \actl f)^\dagger
	= S(P_\mu)^\dagger \actl f^\dagger
\end{align}
for any $f \in \mathbb{A}_\ell$, where $S$ is the antipode of $\Tran{d}_\ell$
\begin{align*}
	P_\mu \actl (e^{ipx})^\dagger
	&= \big( S(P_\mu)^\dagger \actl e^{ipx} \big)^\dagger
	= \big( (\dminus p)_\mu^\dagger e^{i p x} \big)^\dagger
	= (e^{i p x})^\dagger (\dminus p)_\mu
\end{align*}
which proves that $(e^{ipx})^\dagger$ has eigenvalue $\dminus p$ when acted on by translation generators. It therefore corresponds to $e^{i(\dminus p) x}$. With such an equality proven, one can write
\begin{align*}
	\mathscr{I}(f^\dagger)(x)
	&= \int \Haar{p} \mathscr{I} \big( \overline{f}(p) (e^{i p x})^\dagger  \big)\\
	&= \int \Haar{p} \mathscr{I}(\dminus p) \overline{f}(p) e^{i (\dminus p) x} \\
	&= \int \Haar{p} \mathscr{I}(p)^{-1} \overline{f}(p) e^{i (\dminus p) x} \\
	&= (\mathscr{I}^{-1}(f))^\dagger(x).
\end{align*}

\subsection{Proof that \tops{$\omega$}{omega} is a KMS weight}
\label{sapx:KMS_ome}
\paragraph{}
We want to show that $\omega(f) = \int \tdl{x} f(x)$ is a KMS weight for the $1$-parameter group of automorphism $\{\mathscr{I}^{it}\}_{t \in \Real}$ (also called $1$-parameter group of representation or simply $1$-parameter group).

First, we need to show that $\{\mathscr{I}^{it}\}_{t \in \Real}$ is indeed a norm-continuous $1$-parameter group. It requires the following conditions \cite{Kustermans_1997} (see Terminology 1.3):
\begin{enumerate}[label=(\roman*)]
	\item \label{it:opf_tr}
	For every $s, t \in \Real$, we have that $\mathscr{I}^{i(s+t)} = \mathscr{I}^{is} \mathscr{I}^{it}$,
	\item \label{it:opf_id}
	$\mathscr{I}^{i0} = \id$,
	\item \label{it:opf_n}
	For every $t \in \Real$, we have that $\Vert \mathscr{I}^{it} \Vert \leqslant 1$,
	\item \label{it:opf_nc}
	For any $f \in \mathbb{A}$, the map $t \mapsto \mathscr{I}^{it}(f)$ is norm-continuous.
\end{enumerate}
The conditions \ref{it:opf_tr} and \ref{it:opf_id} are directly obtained from the fact that this family is defined through an exponentiation. It remains to check condition \ref{it:opf_n} and \ref{it:opf_nc}.

From appendix \ref{sapx:KMS_ro}, $\mathscr{I}$ is a regular automorphism, so that for any $t \in \Real$, $f \in \mathbb{A}$, $(\mathscr{I}^t(f))^\dagger = \mathscr{I}^{-t}(f^\dagger)$. This implies that $\mathscr{I}^{it}$ is self-adjoint for ${}^\dagger$, \ie $(\mathscr{I}^{it}(f))^\dagger = \mathscr{I}^{it}(f^\dagger)$. This allows us to write
\begin{align*}
	\int \tdl{x} \mathscr{I}^{it}(f)(x) \big(\mathscr{I}^{it}(f)\big)^\dagger(x)
	&= \int \tdl{x} \mathscr{I}^{it}(f)(x) \mathscr{I}^{it}(f^\dagger)(x) \\
	&= (2\pi)^{d+1} \int \Haar{p} \Haar{q} \mathscr{I}^{it}(p) f(p) \mathscr{I}^{it}(q) \overline{f}(q) \delta(p \dplus q) \\
	&= (2\pi)^{d+1} \int \Haar{p} \Haar{q} \mathscr{I}^{it}(p \dplus q) f(p) \overline{f}(q) \delta(p \dplus q) \\
	&= (2\pi)^{d+1} \int \Haar{p} \Haar{q} f(p) \overline{f}(q) \delta(p \dplus q) \\
	&= \int \tdl{x} f(x) f^\dagger(x),
\end{align*}
which can be written as $\Vert \mathscr{I}^{it}(f) \Vert = \Vert f \Vert$ and so that \ref{it:opf_n} is fulfilled. This proves that $\Vert \mathscr{I}^{it} \Vert = 1$. As a side remark, one can notice that this property directly follows from the translation invariance of the Lebesgue measure $\td x$, made explicit in the momentum space by the use of the delta function $\delta(p \dplus q)$.

Finally, the condition \ref{it:opf_nc} is satisfied thanks to the norm-continuity of the exponential map.

Note that we also just showed that $(\Real, \mathbb{A}_\ell, \{\mathscr{I}^{it}\}_{t \in \Real})$ forms a $C^*$-dynamical system (see for example Definition 3.8.1 of \cite{Landsman_1998}).

\paragraph{}
There remains two conditions to show that $\omega$ is a KMS weight with modular group $\{\mathscr{I}^{it}\}_{t \in \Real}$, which corresponds to the Definition 2.8 of a KMS weight of \cite{Kustermans_1997}:
\begin{enumerate}[label=(\roman*), resume]
	\item \label{it:KMS_inv}
	For every $t \in \Real$, $\omega(\mathscr{I}^{it}) = \omega$,
	\item \label{it:KMS_cyc}
	For any $f$ in the domain of $\mathscr{I}^{-1/2}$, one has $\omega(f^\dagger f) = \omega( \mathscr{I}^{-1/2}(f) \, (\mathscr{I}^{-1/2}(f))^\dagger)$.
\end{enumerate}

The property \ref{it:KMS_inv} follows from a computation in momentum space. Explicitly, for any $f \in \mathbb{A}$,
\begin{align*}
	\int \tdl{x} \mathscr{I}^{it}(f)(x)
	&= (2\pi)^{d+1} \int \Haar{p} \mathscr{I}^{it}(p) f(p) \delta(p)
	= (2\pi)^{d+1} \int \Haar{p} f(p) \delta(p)
	= \int \tdl{x} f(x).
\end{align*}
Note that it could have been proved using $\Vert \mathscr{I}^{it} \Vert = 1$ directly.

The last condition \ref{it:KMS_cyc} follows from the twisted cyclicity property of $\omega$ \eqref{eq:Md_cyc}. Indeed, one has
\begin{align*}
	\int \tdl{x} \mathscr{I}^{-1/2}(f)(x) \big(\mathscr{I}^{-1/2}(f)\big)^\dagger(x)
	&= \int \tdl{x} \mathscr{I}^{-1/2}(f)(x) \mathscr{I}^{1/2}(f^\dagger)(x) \\
	&= \int \tdl{x} \mathscr{I}^{-1/2}(f^\dagger)(x) \mathscr{I}^{-1/2}(f)(x) \\
	&= \int \tdl{x} \mathscr{I}^{-1/2}(f^\dagger f)(x) \\
	&= \int \tdl{x} f^\dagger(x) f(x),
\end{align*}
where we have used that $\mathscr{I}$ is a regular automorphism, \ie $(\mathscr{I}(f))^\dagger = \mathscr{I}^{-1}(f^\dagger)$ and $\mathscr{I}(fg) = \mathscr{I}(f) \mathscr{I}(g)$, and the last equation can be proven similarly as the proof used for \ref{it:opf_n}\footnote{
	To make it fully explicit, we have used for the proof of \ref{it:opf_n}, \ref{it:KMS_inv} and \ref{it:KMS_cyc} the same property that, for any $z \in \Cpx$, $\omega(\mathscr{I}^z) = \omega$, which can be proved by going in momentum space. Still, we wanted to highlight the importance of the translation invariance of $\td x$, as done in the proof of \ref{it:opf_n}.
}.



\end{document}